\documentclass[aps,prl,twocolumn,showpacs,superscriptaddress,amsmath]{revtex4}  

\usepackage{graphicx}
\usepackage{dcolumn}
\usepackage{bm}
\usepackage{slashed}
\declareslashed{}{/}{.05}{0}{\ell}

\newcommand{\ppb}{\mbox{\ensuremath{p\bar p}}}

\newcommand{\ttb}{\mbox{\ensuremath{t\bar t}}}

\newcommand{\invfb}{fb$^{-1}$}
\newcommand{\Gev}{\text{GeV}}
\newcommand{\Gc}{\Gev}
\newcommand{\Gcc}{\Gev}

\newcommand{\MH}{\ensuremath{M_H}}
\newcommand{\MW}{\ensuremath{M_W}}

\newcommand{\met}{\mbox{\ensuremath{\not\!\!E_T}}}

\newcommand{\pt}{\mbox{\ensuremath{p_{T}}}}
\newcommand{\etadet}{\mbox{\ensuremath{\vert\eta_{\text{det}}\vert}}}

\newcommand{\HWW}{\mbox{\ensuremath{H\rightarrow WW}}}
\newcommand{\HWpWm}{\mbox{\ensuremath{H\rightarrow W^{+}W^{-}}}}
\newcommand{\WH}{\mbox{\ensuremath{WH\rightarrow \ell\nu bb}}}
\newcommand{\WHbbb}{\mbox{\ensuremath{WH\rightarrow \ell\nu b\bar b}}}

\newcommand{\HZZ}{\mbox{\ensuremath{H\rightarrow ZZ}}}
\newcommand{\HWWllnunu}{\mbox{\ensuremath{\HWpWm \rightarrow \bar\ell\nu\ell'\bar\nu'}}}
\newcommand{\nbHWWllnunu}{\mbox{\ensuremath{WW \rightarrow \ell\nu\ell'\nu'}}}

\newcommand{\HWWlnuqq}{\mbox{\ensuremath{\HWW \rightarrow \ell\nu q'q}}}
\newcommand{\HWWlnuqbq}{\mbox{\ensuremath{\HWW \rightarrow \ell\nu q'\bar q}}}
\newcommand{\HWWtnuqq}{\mbox{\ensuremath{\HWW \rightarrow \tau\nu q'q}}}

\newcommand{\notell}{\ensuremath{\slashed\ell}}
\newcommand{\HXlqq}{\mbox{\ensuremath{H+X\to\ell+\notell /\nu +qq}}}
\newcommand{\HZZlqq}{\mbox{\ensuremath{\HZZ \rightarrow \ell\notell qq}}}

\begin{document}

\hspace{5.2in} \mbox{FERMILAB-PUB-11/018-E}”
\title{\boldmath 
Search for the Standard Model Higgs Boson in the 
$H\to WW\to\ell\nu q'\bar q$ Decay Channel}

%
\affiliation{Universidad de Buenos Aires, Buenos Aires, Argentina}
\affiliation{LAFEX, Centro Brasileiro de Pesquisas F{\'\i}sicas, Rio de Janeiro, Brazil}
\affiliation{Universidade do Estado do Rio de Janeiro, Rio de Janeiro, Brazil}
\affiliation{Universidade Federal do ABC, Santo Andr\'e, Brazil}
\affiliation{Instituto de F\'{\i}sica Te\'orica, Universidade Estadual Paulista, S\~ao Paulo, Brazil}
\affiliation{Simon Fraser University, Vancouver, British Columbia, and York University, Toronto, Ontario, Canada}
\affiliation{University of Science and Technology of China, Hefei, People's Republic of China}
\affiliation{Universidad de los Andes, Bogot\'{a}, Colombia}
\affiliation{Charles University, Faculty of Mathematics and Physics, Center for Particle Physics, Prague, Czech Republic}
\affiliation{Czech Technical University in Prague, Prague, Czech Republic}
\affiliation{Center for Particle Physics, Institute of Physics, Academy of Sciences of the Czech Republic, Prague, Czech Republic}
\affiliation{Universidad San Francisco de Quito, Quito, Ecuador}
\affiliation{LPC, Universit\'e Blaise Pascal, CNRS/IN2P3, Clermont, France}
\affiliation{LPSC, Universit\'e Joseph Fourier Grenoble 1, CNRS/IN2P3, Institut National Polytechnique de Grenoble, Grenoble, France}
\affiliation{CPPM, Aix-Marseille Universit\'e, CNRS/IN2P3, Marseille, France}
\affiliation{LAL, Universit\'e Paris-Sud, CNRS/IN2P3, Orsay, France}
\affiliation{LPNHE, Universit\'es Paris VI and VII, CNRS/IN2P3, Paris, France}
\affiliation{CEA, Irfu, SPP, Saclay, France}
\affiliation{IPHC, Universit\'e de Strasbourg, CNRS/IN2P3, Strasbourg, France}
\affiliation{IPNL, Universit\'e Lyon 1, CNRS/IN2P3, Villeurbanne, France and Universit\'e de Lyon, Lyon, France}
\affiliation{III. Physikalisches Institut A, RWTH Aachen University, Aachen, Germany}
\affiliation{Physikalisches Institut, Universit{\"a}t Freiburg, Freiburg, Germany}
\affiliation{II. Physikalisches Institut, Georg-August-Universit{\"a}t G\"ottingen, G\"ottingen, Germany}
\affiliation{Institut f{\"u}r Physik, Universit{\"a}t Mainz, Mainz, Germany}
\affiliation{Ludwig-Maximilians-Universit{\"a}t M{\"u}nchen, M{\"u}nchen, Germany}
\affiliation{Fachbereich Physik, Bergische Universit{\"a}t Wuppertal, Wuppertal, Germany}
\affiliation{Panjab University, Chandigarh, India}
\affiliation{Delhi University, Delhi, India}
\affiliation{Tata Institute of Fundamental Research, Mumbai, India}
\affiliation{University College Dublin, Dublin, Ireland}
\affiliation{Korea Detector Laboratory, Korea University, Seoul, Korea}
\affiliation{CINVESTAV, Mexico City, Mexico}
\affiliation{FOM-Institute NIKHEF and University of Amsterdam/NIKHEF, Amsterdam, The Netherlands}
\affiliation{Radboud University Nijmegen/NIKHEF, Nijmegen, The Netherlands}
\affiliation{Joint Institute for Nuclear Research, Dubna, Russia}
\affiliation{Institute for Theoretical and Experimental Physics, Moscow, Russia}
\affiliation{Moscow State University, Moscow, Russia}
\affiliation{Institute for High Energy Physics, Protvino, Russia}
\affiliation{Petersburg Nuclear Physics Institute, St. Petersburg, Russia}
\affiliation{Stockholm University, Stockholm and Uppsala University, Uppsala, Sweden }
\affiliation{Lancaster University, Lancaster LA1 4YB, United Kingdom}
\affiliation{Imperial College London, London SW7 2AZ, United Kingdom}
\affiliation{The University of Manchester, Manchester M13 9PL, United Kingdom}
\affiliation{University of Arizona, Tucson, Arizona 85721, USA}
\affiliation{University of California Riverside, Riverside, California 92521, USA}
\affiliation{Florida State University, Tallahassee, Florida 32306, USA}
\affiliation{Fermi National Accelerator Laboratory, Batavia, Illinois 60510, USA}
\affiliation{University of Illinois at Chicago, Chicago, Illinois 60607, USA}
\affiliation{Northern Illinois University, DeKalb, Illinois 60115, USA}
\affiliation{Northwestern University, Evanston, Illinois 60208, USA}
\affiliation{Indiana University, Bloomington, Indiana 47405, USA}
\affiliation{Purdue University Calumet, Hammond, Indiana 46323, USA}
\affiliation{University of Notre Dame, Notre Dame, Indiana 46556, USA}
\affiliation{Iowa State University, Ames, Iowa 50011, USA}
\affiliation{University of Kansas, Lawrence, Kansas 66045, USA}
\affiliation{Kansas State University, Manhattan, Kansas 66506, USA}
\affiliation{Louisiana Tech University, Ruston, Louisiana 71272, USA}
\affiliation{Boston University, Boston, Massachusetts 02215, USA}
\affiliation{Northeastern University, Boston, Massachusetts 02115, USA}
\affiliation{University of Michigan, Ann Arbor, Michigan 48109, USA}
\affiliation{Michigan State University, East Lansing, Michigan 48824, USA}
\affiliation{University of Mississippi, University, Mississippi 38677, USA}
\affiliation{University of Nebraska, Lincoln, Nebraska 68588, USA}
\affiliation{Rutgers University, Piscataway, New Jersey 08855, USA}
\affiliation{Princeton University, Princeton, New Jersey 08544, USA}
\affiliation{State University of New York, Buffalo, New York 14260, USA}
\affiliation{Columbia University, New York, New York 10027, USA}
\affiliation{University of Rochester, Rochester, New York 14627, USA}
\affiliation{State University of New York, Stony Brook, New York 11794, USA}
\affiliation{Brookhaven National Laboratory, Upton, New York 11973, USA}
\affiliation{Langston University, Langston, Oklahoma 73050, USA}
\affiliation{University of Oklahoma, Norman, Oklahoma 73019, USA}
\affiliation{Oklahoma State University, Stillwater, Oklahoma 74078, USA}
\affiliation{Brown University, Providence, Rhode Island 02912, USA}
\affiliation{University of Texas, Arlington, Texas 76019, USA}
\affiliation{Southern Methodist University, Dallas, Texas 75275, USA}
\affiliation{Rice University, Houston, Texas 77005, USA}
\affiliation{University of Virginia, Charlottesville, Virginia 22901, USA}
\affiliation{University of Washington, Seattle, Washington 98195, USA}
\author{V.M.~Abazov} \affiliation{Joint Institute for Nuclear Research, Dubna, Russia}
\author{B.~Abbott} \affiliation{University of Oklahoma, Norman, Oklahoma 73019, USA}
\author{B.S.~Acharya} \affiliation{Tata Institute of Fundamental Research, Mumbai, India}
\author{M.~Adams} \affiliation{University of Illinois at Chicago, Chicago, Illinois 60607, USA}
\author{T.~Adams} \affiliation{Florida State University, Tallahassee, Florida 32306, USA}
\author{G.D.~Alexeev} \affiliation{Joint Institute for Nuclear Research, Dubna, Russia}
\author{G.~Alkhazov} \affiliation{Petersburg Nuclear Physics Institute, St. Petersburg, Russia}
\author{A.~Alton$^{a}$} \affiliation{University of Michigan, Ann Arbor, Michigan 48109, USA}
\author{G.~Alverson} \affiliation{Northeastern University, Boston, Massachusetts 02115, USA}
\author{G.A.~Alves} \affiliation{LAFEX, Centro Brasileiro de Pesquisas F{\'\i}sicas, Rio de Janeiro, Brazil}
% ...
%
%
\collaboration{The D0 Collaboration\footnote{with visitors from
$^{a}$Augustana College, Sioux Falls, SD, USA,
$^{b}$The University of Liverpool, Liverpool, UK,
$^{c}$SLAC, Menlo Park, CA, USA,
$^{d}$ICREA/IFAE, Barcelona, Spain,
$^{e}$Centro de Investigacion en Computacion - IPN, Mexico City, Mexico,
$^{f}$ECFM, Universidad Autonoma de Sinaloa, Culiac\'an, Mexico,
and 
$^{g}$Universit{\"a}t Bern, Bern, Switzerland.%
}} \noaffiliation
\vskip 0.25cm

\date{January 29, 2011}
          
\begin{abstract}

We present a search for the standard model Higgs boson (H) in \ppb\ 
collisions at $\sqrt{s}=1.96$~TeV in events containing a charged 
lepton ($\ell$), missing transverse energy, and at least two jets, 
using 5.4~\invfb\ of integrated luminosity recorded with the D0 detector 
at the Fermilab Tevatron Collider.  This analysis is sensitive primarily 
to Higgs bosons produced through the fusion of two gluons or two 
electroweak bosons, with subsequent decay \HWWlnuqbq , where $\ell$ 
is an electron or muon. The search is also sensitive to 
contributions from other production channels, such as \WHbbb .
In the absence of signal, we set limits at 
the 95\% C.L.\ on the cross section for H production 
$\sigma(p\bar p\rightarrow H+X)$ in these final states.
For a mass of $\MH=160$\,\Gcc , the limit is a factor of 3.9 
larger than the cross section in the standard model, and consistent 
with expectation.

\end{abstract}


\maketitle


The Higgs 
mechanism~\cite{PhysRevLett.13.508,Englert:1964et,Guralnik:1964eu,PhysRev.145.1156} 
accommodates the observed breaking of electroweak (EW) symmetry
in the standard model (SM).  In addition to generating masses 
for the EW $W$ and $Z$ bosons, as well as for fermions, the theory
predicts a new scalar Higgs boson ($H$) with well-determined couplings, but
unknown mass ($M_H$). Confirmation of the existence and properties of the $H$ 
boson would be a key step in elucidating the origins of EW 
symmetry breaking. For a Higgs boson with mass $M_H \gtrsim 135$\,\Gcc , 
the dominant decay mode is \HWpWm , where at least one $W$ boson
must be virtual when $\MH < 2\MW$.  
Previous searches~\cite{D0wwllnn,CDFwwllnn,HiggsCombo} 
for this process were based on events with two charged leptons ($\ell$) 
and large missing transverse energy (\met ) from the decay \HWWllnunu\
($\ell=e,\mu$).   This Letter presents the first search 
for production of Higgs bosons with subsequent decay to $WW$ having
only one charged lepton in the final state.
The data sample is 5.4~\invfb\ of integrated luminosity in 
\ppb\ collisions at $\sqrt{s}=1.96$\,TeV recorded with the D0 detector
at the Fermilab Tevatron Collider.  
The largest SM contributions to the inclusive cross section for 
producing $H$ bosons in \ppb\ collisions are from mechanisms
involving the fusion of two gluons or two weak vector bosons
into an $H$ boson, and associated production of $H$ and a weak vector boson
($V=W\text{ or }Z$).    In the following we will not distinguish between 
particles and antiparticles.
The most striking signatures from $H\to VV$ decays are in the purely
leptonic final states, but these account for only 5\% of such decays. 
Final states containing a single charged lepton 
have larger backgrounds, but their branching fractions are a 
factor of $\approx 6$ 
larger than for the all-leptonic modes.  

A recent calculation of the differential width for \HWWlnuqq\
decays~\cite{HiggsWidth} supports the important role of these mixed 
modes for characterizing a potential SM Higgs-boson signal.
Our analysis is most sensitive to final-state 
topologies with a single lepton ($e$ or $\mu$), two or more jets, 
and \met , arising from \HWWlnuqq\ decays.
For $\MH\lesssim 140$\,\Gcc , significant sensitivity 
is gained from \WH , where we do not attempt to identify the $b$ quark flavor.
Smaller signal contributions from \HZZlqq , where \notell\ represents 
an unidentified lepton, and \HWWtnuqq\ with $\tau\to\ell\nu\nu$ 
are also included. For $\MH \ge 160$\,\Gcc , assuming
that the observed \met\ is due to the neutrino from the decay of a $W$ boson,
it is possible to reconstruct the longitudinal momentum $(p_z^\nu)$ 
up to a twofold ambiguity, and 
thereby extract the mass of the $H$ decaying to $WW$~\cite{GunionSoldate}.
We choose the solution with smallest $|\text{Re}(p_z^\nu)|$ 
to calculate \MH .
The primary backgrounds are from $V$+jets, top quark, diboson 
production, and multijet (MJ) events containing 
a lepton or lepton-like signature
with \met\ generally arising from mismeasurement of jet energies.


The D0 detector~\cite{Dzero} consists of tracking, calorimetric and muon 
detectors. Charged particle tracks are reconstructed using silicon microstrip 
(SMT) detectors and a scintillating fiber tracker, within a 2\,T solenoid.  
Three uranium/liquid-argon calorimeters measure particle energies that
are reconstructed into hadronic jets using an iterative midpoint cone 
algorithm with a cone radius of 0.5~\cite{jetalgo}.
Electrons and muons are identified through association of 
charged particle tracks with clusters in the electromagnetic sections of the
calorimeters or with hits in the muon detector, respectively.  
We obtain the \met\ from a vector sum of transverse components of 
calorimeter energy depositions and correct it for identified muons. 
Jet energies are calibrated 
using transverse momentum balance in photon+jet events~\cite{d0incjet}, 
and the correction is propagated to the \met . 
The data are recorded using triggers designed to select single electrons 
or muons  and combination of  electron and jets.  
After imposing data quality requirements, 
the total integrated luminosity is 5.4~\invfb ~\cite{lumi}, where
the first 1.1~\invfb , Run~IIa, precedes an upgrade 
to the SMT and trigger systems. The remaining 4.3~\invfb\ is
denoted as Run~IIb.  The four data sets using $e$ or $\mu$ for 
the two run epochs are analyzed separately and combined in 
the final result.

Background contributions from most SM processes are determined through
Monte Carlo (MC) simulation, while multijet background 
is estimated from data. The dominant background is from
$V$+jets processes, which are generated with {\sc alpgen}~\cite{alpgen}.
The transverse momentum (\pt\ ) spectrum of the $Z$ boson in the MC 
is reweighted to match that observed in data~\cite{Zpt}. The \pt\ spectrum 
of the $W$ boson is reweighted using the same dependence, but
corrected for differences between the \pt\ spectra of $Z$ and $W$ bosons
predicted in next-to-next-to-leading order (NNLO) QCD~\cite{MelPet}.
Backgrounds from \ttb\ and electroweak single top quark production 
are simulated using the {\sc alpgen} and {\sc comphep}~\cite{comphep} 
generators, respectively.  Vector boson pair production and $H$ boson 
signals are generated with {\sc pythia}~\cite{pythia}.  All these simulations 
use CTEQ6L1 parton distribution functions (PDF)~\cite{cteq}.  Both {\sc alpgen}
and {\sc comphep} samples are interfaced with {\sc pythia} for modeling 
of parton evolution and hadronization.

Relative normalizations for the various $V$+jets processes 
are obtained from calculations of cross sections at next-to-leading 
order using {\sc mcfm}~\cite{mcfm}, while the absolute normalization for the 
total $V$+jets background is constrained through a comparison 
to data, following the subtraction of other background 
sources. This increases the normalization for $V$+jets background by
about 2\%, compared with the expectation from {\sc alpgen}
normalized using total cross sections calculated at
NNLO~\cite{WZcross} with the MRST2004 NNLO PDFs~\cite{MRST}.
Cross sections for other SM backgrounds are taken from
Ref.~\cite{xsections}, or calculated with {\sc mcfm}, 
and those for signal are taken from Ref.~\cite{signal}.
The \pt\ spectra for diboson events in
background are corrected to match those of the {\sc mc@nlo} 
generator~\cite{mc@nlo}.
The \pt\ spectra from the contribution of gluon fusion to 
the $H$ boson signal, generated in {\sc pythia}, are modified to match 
those obtained from {\sc sherpa}~\cite{sherpa}.

Signal and background events from MC are passed through a full 
{\sc geant3}-based simulation~\cite{geant} of detector response, then 
processed with the same reconstruction program as used for data. 
Events from randomly selected beam crossings with the same instantaneous 
luminosity profile as data are overlaid on the simulated 
events to model detector noise and contributions from the presence of
additional \ppb\ interactions.  Parameterizations of trigger efficiency 
for leptons are determined using $Z\to\ell\ell$ 
decays~\cite{d0tagprobe}.  
Any remaining differences between data and simulation in the
reconstruction of electrons, muons, and jets are adjusted in simulated
events to match those observed in data and these corrections are
propagated to the MET.


\begin{table*}[ht]
\caption{\label{t:yields}
Number of signal and background events expected after selection 
requirements. The signal sources include gluon-gluon and 
vector-boson fusion, and associated production $WH$.  
The three numbers quoted for the signals 
correspond to $\MH=130, 160, \text{and }190\,\Gev$. 
For backgrounds, ``top'' includes pair and single 
top quark production and ``$VV$'' includes all non-signal diboson 
processes.  The overall background normalization is fixed 
to the data by adjusting the $V$+jets cross sections.} 

\begin{ruledtabular}
\begin{tabular}{lccccccccc}
Channel  &$gg\to H$       &$qq\to qqH$  & $WH$    & $V$+jets & Multijet & Top   & $VV$ & Total Background & Data  \\
Electron & 11.2/46.3/27.8 & 2.1/6.4/4.2 & 7.2/0/0 & 52158    & 11453    & 2433	& 1584 & 67627            & 67627 \\
Muon     & 9.5/34.7/20.4  & 1.5/4.4/2.9 & 5.7/0/0 & 47970    & 2720     & 1598  & 1273 & 53562            & 53562 \\
\end{tabular}
\end{ruledtabular}
\end{table*}

Events are selected to contain candidates 
for $W\to\ell\nu$ decay by requiring $\met > 15$\,GeV and the presence 
of a lepton with $\pt > 15$\,\Gc\ that is isolated relative to jets, namely
located outside jet cones $\Delta R(\ell,j)>0.5$, 
with $(\Delta R)^2=(\phi^\ell-\phi^j)^2 + (\eta^\ell-\eta^j)^2$, 
where $\phi^x$ and $\eta^x$ are the azimuth and pseudorapidity~\cite{eta} of object $x.$
The \ppb\ interaction vertex (PV) position along the beam direction 
($z_{\text{PV}}$) is required to be reconstructed within the longitudinal acceptance 
of the SMT, $\vert z_{\text{PV}}\vert < 60$~cm.
The lepton is required to originate from the PV and to pass 
more restrictive isolation criteria based on tracking information 
and energy deposited near the lepton in the calorimeter.  
Electrons must also satisfy criteria for the spatial distribution of 
the shower, and timing information is used to reject cosmic ray background 
in events with muons.  All lepton selections are described 
in Ref.~\cite{d0sngltop}, except that this analysis requires both 
the scalar sum of 
track \pt\ and calorimeter energy in the vicinity of the muon be 
less than 2.5\,\Gev .
Electrons and muons are required to be located within $\etadet <1.1$
and $<1.6$, respectively, where $\eta_{det}$ is the pseudorapidity 
assuming the object originates from the center of the detector.
To reduce background from $Z\to\ell\ell$, top quark, and diboson events, 
and to assure selected events do not overlap with those used in the
\nbHWWllnunu\ analysis channels, we veto any event containing 
an additional lepton satisfying less stringent identification criteria.  
We also require at least two jets with $|\eta^j|<2.5$ and $p_T>20$\,\Gc\
that contain associated tracks originating from the PV.  The jet \pt\ 
requirement is 23\,\Gc\ when the second-leading jet (ordered in $p_T$) 
has $0.8<|\eta_{det}|<1.5$~\cite{Dzero}.
The leading two jets are used to reconstruct the $W$ boson 
decaying to $q'q$.
To suppress background from MJ events~\cite{d0sngltop0}, we require 
events to have $M_T^W(\Gcc)>40-0.5\times\met$, where $M_T^W$
is the transverse mass~\cite{d0wmass} of the $W$ boson candidate.

To estimate the MJ background, we use data samples orthogonal 
to our signal sample. For the electron channel, we form a ``loose'' 
category for which the selection on a likelihood discriminant 
used to select a ``tight'' electron, based on 
calorimeter and track variables~\cite{d0sngltop0}, is reversed. 
Following the method of Ref.~\cite{ttbarprd04},
the MJ background is evaluated from independently-determined 
probabilities for loose electrons or jets to pass the tight 
signal selections.  For the muon channel, we reverse 
requirements on muon isolation in both the tracking 
detectors and the calorimeters, and subtract contributions 
arising from SM processes containing a true muon from $W$
or $Z$ decay.
The normalization is obtained from fits to both the $V$+jets 
and MJ contributions using observed distributions of 
$p_T^\mu$ and \met . Event yields in data and those expected 
for signal and background are shown in Table~\ref{t:yields}.

\begin{figure}[htb]
\centering
\includegraphics[width=8.5cm]{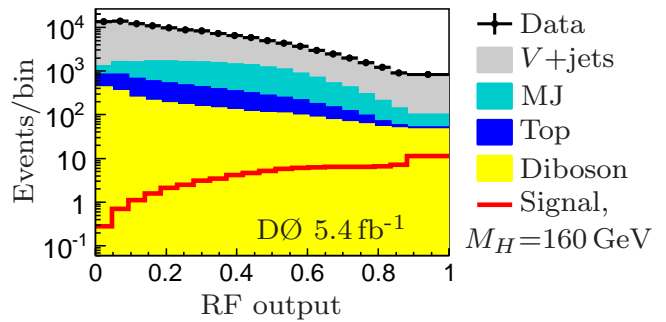}
\caption{\label{dt_output}
The output of RF discriminants for data, different backgrounds, and signal
for $\MH =160\,\Gcc$ for the combined data sets. 
}
\end{figure}

We use a random forest (RF) of 50 decision trees (DT)
to separate signal from background~\cite{Breiman, Narsky}.
Each DT is trained on a randomly selected collection of signal and
background MC events as well as MJ events from data.
The DTs examine a random set of about 30 discriminating variables
formed from particle 4-vectors, angles between 
objects, and combinations of kinematic variables such 
as reconstructed masses and event shapes.  
An RF is trained separately for each data set, 
using signal hypotheses $115 < M_H < 200$\,\Gcc\ in steps of 5\,\Gcc .  
The outputs of the final RF discriminants 
for the four data sets combined, background, and signal for $M_H=160$\,\Gcc\
are shown in Fig.~\ref{dt_output} .
Agreement is observed with expectations from SM background, and
the RF-output distributions are therefore used to set upper
limits on the cross section for SM Higgs production.

Systematic uncertainties affect the normalizations and 
distributions of the final discriminants and are included in
the determination of limits. These
arise from a variety of sources, and their impact
is assessed by changing each input discriminant to the RF by $\pm 1$
standard deviation.  The most significant uncertainties are from
calibration of jet
energies (0.7--6)\%, jet resolution (0.5--3)\%, jet reconstruction 
efficiency (0.5--4)\%, lepton identification and modeling of the 
trigger (4\%), estimation of the multijet background (6.5--26)\%, 
and integrated luminosity (6.1\%).  
Theoretical uncertainties on cross sections for backgrounds are taken 
from Ref.~\cite{xsections,mcfm}.  The uncertainties on cross 
sections for signal are taken from Ref.~\cite{signal}.  Because the 
overall cross section for $V$+jets production is constrained by data, 
the uncertainty on its normalization is anti-correlated with the MJ 
background.  The impacts of theoretical uncertainties on distributions 
of the final discriminants are assessed by varying a common
renormalization and factorization scale, by 
comparing {\sc alpgen} interfaced with {\sc herwig}~\cite{herwig} 
to {\sc alpgen} interfaced with {\sc pythia} for $V$+jets samples,
and by varying PDF parameters using the prescription of Ref.~\cite{cteq} for 
all MC samples.  

\begin{figure}[thb]
\centering
\includegraphics[width=7cm]{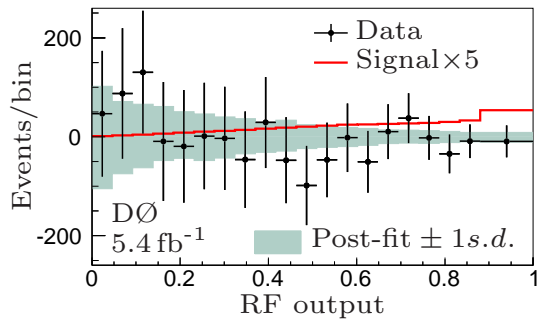}
\caption{\label{dataminusb}
The combined background subtracted data and
one standard deviation (s.d.)~uncertainty on total 
background after applying constraints on systematic uncertainties 
by fitting to data.  The expected SM Higgs signal for $\MH=160$\,\Gcc , 
shown by the line, is scaled up by a factor of 5.
}
\end{figure}

 
Upper limits on the production cross section multiplied by 
branching fractions are determined using the modified frequentist $CL_S$ approach~\cite{cls}.
A test statistic based on the logarithm of the ratio of 
likelihoods (LLR)~\cite{cls}
for the data to represent signal+background and background-only hypotheses
is summed over all bins of the final discriminant in each 
set.    To minimize degradation in sensitivity, 
scaling factors for the systematic uncertainties
are fitted 
to the data by maximizing a likelihood function for both the 
signal+background and background-only hypotheses, with 
the systematic uncertainties constrained through Gaussian priors on their 
probabilities~\cite{wade}.
Correlations among systematic uncertainties in signal and 
background are taken into account in extracting the final results.  
Figure~\ref{dataminusb} shows the combined background-subtracted data  
and the uncertainties on the RF discriminant after they
are fitted to the data.

\begin{table*}
\caption{\label{limits} 
Ratios of the observed and expected exclusion limits
relative to the SM production cross section for 
$\sigma(p\bar p\rightarrow H+X)$ multiplied by
the branching fraction for
\HXlqq\ at the 95\% C.L.\ as a function of $M_H$. 
}
\begin{ruledtabular}

\begin{tabular}{lcccccccccccccccccc}

$M_H$ (\Gcc) & 115 & 120 & 125 & 130 & 135 & 140 & 145 & 150 & 155 & 160 & 165 & 170 & 175 & 180 & 185 & 190 & 195 & 200 \\ \hline
Observed & 28.5 & 20.4 & 32.8 & 36.6 & 33.0 & 33.7 & 23.1 & 17.1 & 8.3 & 3.9 & 5.2 & 5.6 & 8.2 & 7.1 & 12.0 & 10.6 & 10.0 & 10.4 \\ 
Expected & 19.5 & 23.4 & 26.4 & 28.4 & 25.7 & 19.7 & 13.7 & 10.4 & 8.0 & 5.0 & 5.1 & 5.9 & 6.7 & 8.0 & 9.6 & 10.7 & 11.2 & 12.1 \\ 
\end{tabular}

\end{ruledtabular}
\end{table*}

\begin{figure}[htbp]
\centering
\includegraphics[width=8.5cm]{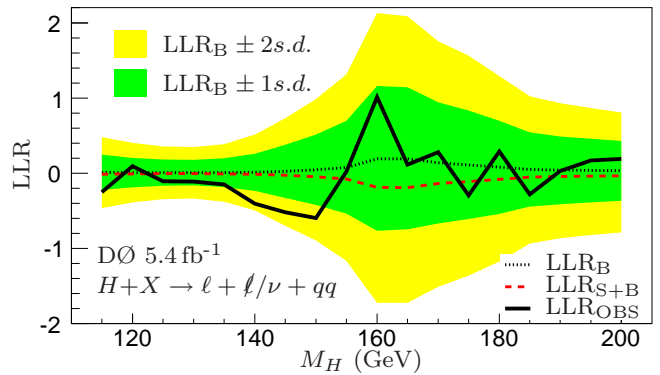}
\caption{\label{limllrs}
The observed LLRs for the combined data sets are given by 
the solid line, the expected LLRs for the background-only and 
signal+background hypotheses 
are shown as dots and dashes, respectively, and the dark 
and light-shaded areas correspond to one and two standard deviations (s.d.)
around the expected LLRs for the background-only hypothesis. 
Negative values of LLR$_{\text{OBS}}$ represent signal-like fluctuations 
in the data.
}
\end{figure}

The resulting limits on standard model Higgs boson production are 
given in Table~\ref{limits}. The LLR$_{\text{OBS}}$
values shown in Fig.~\ref{limllrs} as functions of \MH\ are 
within $\sim1.5$ standard deviations of the expected median for 
LLR$_{\text{B}}$, the background-only hypothesis, as calculated from
statistical fluctuations and systematic uncertainties.

In conclusion, we have determined the first limits on standard model 
Higgs boson production examining decays of the Higgs boson
to two vector bosons, one of which decays leptonically and the 
other into a pair of quarks.
For $\MH =160$\,\Gcc, the observed and expected 95\% C.L.\
upper limits on the combined cross section for Higgs production, 
multiplied by the branching fraction for \HXlqq ,  are factors of 3.9 
and 5.0 larger than the SM cross section, respectively.

Supplementary material, including a list of variables used
in the RF, samples of input distributions, and a table of 
systematic uncertainties, is available from~\cite{appendix}.

%
We thank the staffs at Fermilab and collaborating institutions,
and acknowledge support from the
DOE and NSF (USA);
CEA and CNRS/IN2P3 (France);
FASI, Rosatom and RFBR (Russia);
CNPq, FAPERJ, FAPESP and FUNDUNESP (Brazil);
DAE and DST (India);
Colciencias (Colombia);
CONACyT (Mexico);
KRF and KOSEF (Korea);
CONICET and UBACyT (Argentina);
FOM (The Netherlands);
STFC and the Royal Society (United Kingdom);
MSMT and GACR (Czech Republic);
CRC Program and NSERC (Canada);
BMBF and DFG (Germany);
SFI (Ireland);
The Swedish Research Council (Sweden);
and
CAS and CNSF (China).

\onecolumngrid
\newpage

\begin{table*}
\begin{center}
\textbf{\Large{Supplementary Material}}
\end{center}
\end{table*}

\begin{table*}
\caption{Variables used in the random forest 
classifiers, given in categories 
of object kinematics, angular variables, and event kinematics.  
The symbol ($^\dag$) denotes a measurement in the rest frame 
of the reconstructed $H$ boson.  A subset of these is used for each channel
and data-taking period.}
\begin{ruledtabular}
\begin{tabular}{lll} \\
\textbf{Object Kinematics} & & \mbox{}\\ 
Lepton energy & $E^\ell$ \\
Lepton/jet transverse momentum & $p_T^\ell, p_T^{j_1}, p_T^{j_2}$ & \mbox{} \\ \\ \hline \\
\textbf{Angular Variables} & & \mbox{}\\
Azimuthal angles & $\Delta\phi(j_1,j_2), \Delta\phi(\ell,\met ),$ & \mbox{} \\
                 & $\text{min}[\Delta\phi(\met ,j_1),\Delta\phi(\met ,j_2)],
                    \text{min}[\Delta\phi(\ell,j_1),\Delta\phi(\ell,j_2)]$ & \mbox{} \\
                 & $\Delta\phi(\text{bisector of dijet pair},\met )$ & \mbox{} \\  
Polar angles & $\Delta\eta(j_1,j_2)$  & \mbox{} \\ 
$\eta-\phi$ combinations &  $\Delta R(j_1,j_2), 
              \text{max}[\Delta R(\ell ,j_1),\Delta R(\ell ,j_2)], 
              \text{min}[\Delta R(\met ,j_1),\Delta R(\met ,j_2)]$  & \mbox{} \\ 
3D angles    & $\text{angle}(j_1,j_2)^\dag, \text{angle}(j_1, \text{reconstructed}~W\to\ell\nu)^\dag, 
              \text{angle}(\ell ,\text{dijet system})$  & \mbox{} \\  \\ \hline \\
\textbf{Event Kinematics} & & \\
Velocity         & $\beta(\text{dijet system})$ & \\ 
Transverse momentum sums 
                 & $\met, \pt(\text{dijet system}), \pt(\text{leptonic}~W)$ & \\
                 & $H_T(j_1,j_2)$, $H_T(\text{all jets})$, 
                    $H_T(j_1,j_2,\ell )$, $H_T(j_1,j_2,\ell , \met )$ & \\  
Topological variables$^a$  & $\text{aplanarity}(j_1,j_2,\ell,\nu)$ & \\
                 & $\text{centrality}(\ell,j_1,j_2)$, 
                    $\text{centrality}(\ell,\text{all jets})$ & \\
                 & $\text{sphericity}(\ell ,j_1, j_2)$, 
                    $\text{sphericity}(\ell ,j_1, j_2,\met)$ & \\   
``N''-body masses    &  $ \text{dijet mass}: M(j_1,j_2)$, $M(j_1,j_2,\ell)$, $M(WW)$, 
                    WW~\text{``visible mass''}: $M(j_1,j_2,\ell,\met )$ & \\   
Transverse masses& $M_T(j_1,j_2)$, $M_T(W\to\ell\nu)$,  $M_T(WW)$ &\\  
Ratio of jet energies  & ${E^{j_2}/E^{j_1}}^\dag$ & \\  
$KT^{\text{max/min}}$         
         & $\Delta R(j_1,j_2)\cdot E_T(j_{1/2})/(E_T^\ell + \met )$ &\\  
Other combinations 
                 & scalar product of \met\ and angular bisector of dijet pair & \\
                 & $\Delta\phi(W^{\ell},{\rm bisector~of~dijet~system})$ & \\
           & Magnitude of $\pt^{j_1}$ perpendicular to dijet system & \\    
Scaled \met      & $\displaystyle\sum_{\text{all jets}}
            [\sqrt{E^{j_i}}\sin(\theta^{j_i})\cos(\Delta\phi(j_i,\met ))]^2$ & \\ \\
\end{tabular}
\end{ruledtabular}

\begin{flushleft}
$^a${\footnotesize For descriptions of the topological variables
see: V.D. Barger and R.J.N. Phillips, Collider Physics, 
Addison-Wesley, Reading, MA, 1987.}\\
\end{flushleft}
\end{table*}

\clearpage

\begin{figure*}[htbp]
\centering
\includegraphics{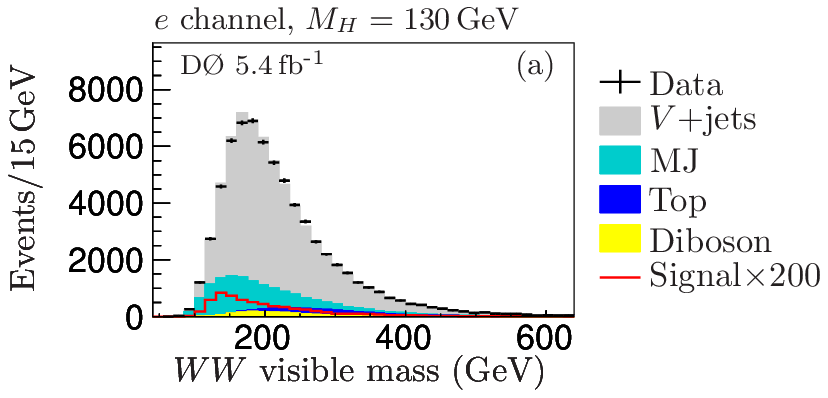} \hfill
\includegraphics{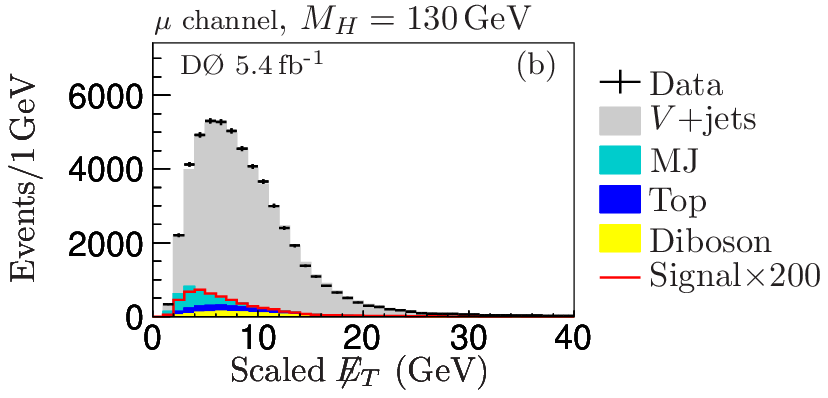} \\ \medskip
\includegraphics{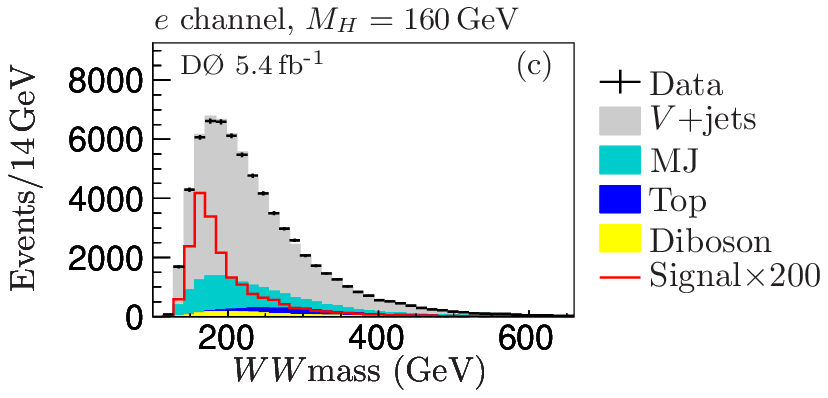}\hfill
\includegraphics{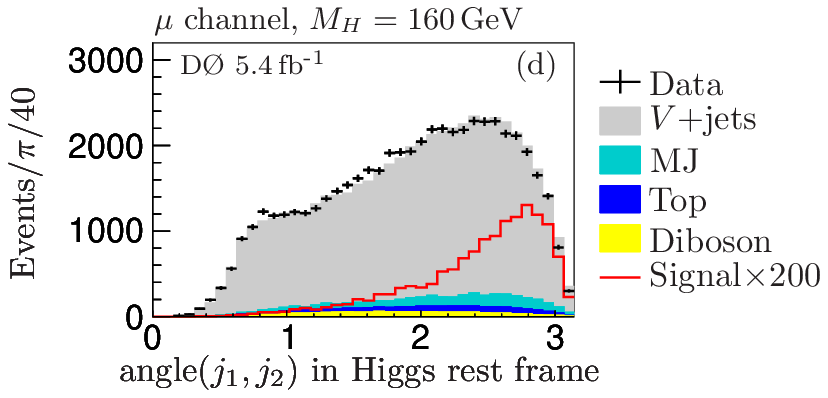} \\ \medskip
\includegraphics{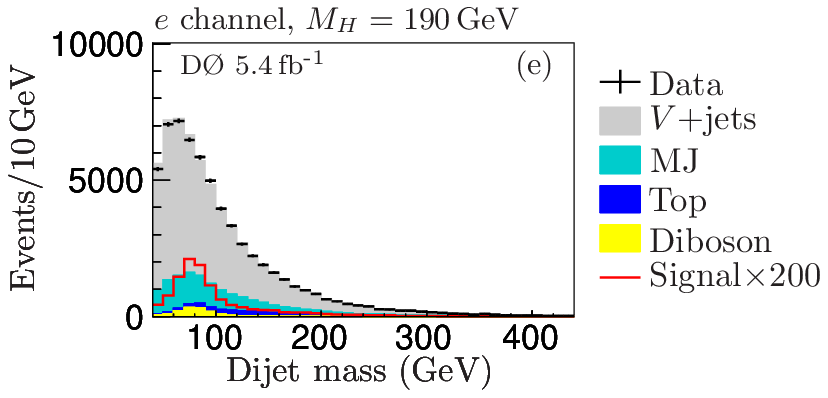}\hfill
\includegraphics{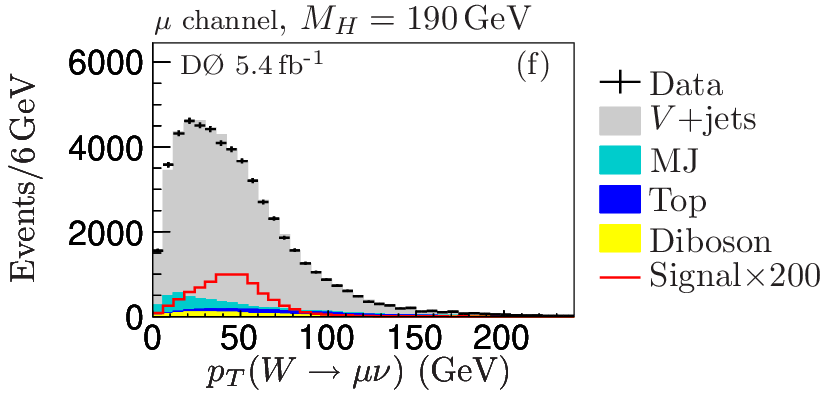}
\caption{Sample of DT inputs for (a,c,e) electron and (b,d,f)
muon channels.  The data are shown as points with uncertainties. 
The background contributions are shown as histograms, 
with sources indicated in the legend.  
``$V$+jets'' includes $(W/Z)$+$(u,d,s,c,b,g)$ jets, MJ is the multijet 
background, ``top'' includes pair and single top quark production, 
diboson includes $WW$, $WZ$, and $ZZ$ processes.
The distributions for signal are multiplied by a factor of 200.
}
\end{figure*}

\begin{table*}
\caption{\label{table:systematics}Systematic uncertainties for the 
electron and muon channels. Signal uncertainties are shown for 
$M_H=160$\,\Gcc\ for all channels except for $WH$, 
shown for $M_H=115$\,\Gcc .  Those affecting the shape of 
the RF discriminant are indicated with ``Y.''
Uncertainties are listed as relative changes in normalization, 
in percent, except for those also marked by ``S,'' where the 
the overall normalization is constant, and the value given
denotes the maximum percentage change from nominal in any region of the
distribution.}
\begin{ruledtabular}

\begin{tabular}{llccccccl}

Contribution & Shape & $W$+jets & $Z$+jets & top & diboson & $gg\to H$ & $qq\to qqH$ & $WH$ \\ \hline
Jet energy scale & Y & $\binom{+6.7}{-5.4}^S$ & $<0.1$ & $\pm$0.7 & $\pm$3.3 & $\binom{+5.7}{-4.0}$ & $\pm$1.5 &$\binom{+2.7}{-2.3}$  \\ 
Jet identification & Y & $\pm 6.6^S$ & $<0.1$ & $\pm$0.5 & $\pm$3.8  & $\pm$1.0 & $\pm$1.1 & $\pm$1.0 \\ 
Jet resolution & Y & $\binom{+6.6}{-4.1}^S$ & $<0.1$ & $\pm$0.5 & $\binom{+1.0}{-0.5}$ & $\binom{+3.0}{-0.5}$ & $\pm 0.8$ & $\pm 1.0$ \\ 
Association of jets with PV & Y & $\pm 3.2^S$ & $\pm 1.3^S$ & $\pm$1.2 & $\pm$3.2 & $\pm$2.9 & $\pm$2.4 & $\binom{+0.9}{-0.2}$ \\ 
Luminosity & N & n/a & n/a & $\pm$6.1 & $\pm$6.1 & $\pm$6.1 & $\pm$6.1 &  $\pm$6.1 \\ 
Muon trigger  & Y & $\pm 0.4^S$ & $<0.1$ & $<0.1$ & $<0.1$ & $<0.1$ & $<0.1$ &  $<0.1$ \\ 
Electron identification & N & $\pm$4.0  & $\pm$4.0  & $\pm$4.0  & $\pm$4.0  & $\pm$4.0  & $\pm$4.0  & $\pm$4.0 \\ 
Muon identification  & N & $\pm$4.0  & $\pm$4.0  & $\pm$4.0  & $\pm$4.0  & $\pm$4.0  & $\pm$4.0  & $\pm$4.0  \\ 
ALPGEN tuning & Y & $\pm 1.1^S$ & $\pm 0.3^S$ & n/a & n/a & n/a & n/a & n/a \\ 
Cross Section & N & $\pm$6 & $\pm$6 &  $\pm$10 & $\pm$7 & $\pm$10 & $\pm$10 & $\pm$6 \\ 
Heavy-flavor fraction  & Y & $\pm$20 & $\pm$20 & n/a & n/a & n/a & n/a & n/a  \\ 
PDF & Y & $\pm 2.0^S$ & $\pm 0.7^S$ & $<0.1^S$ & $<0.1^S$ & $<0.1^S$ & $<0.1^S$ & $<0.1^S$ \\ 
 &  &  &  &  &  &  &  &  \\ 
 &  & \multicolumn{ 3}{c}{Electron channel} & \multicolumn{ 3}{c}{Muon channel} &  \\ 
Multijet Background & Y  & \multicolumn{ 3}{c}{$\pm$6.5} & \multicolumn{ 3}{c}{$\pm$26} &  \\ 
\end{tabular}
\end{ruledtabular}
\end{table*}

\bigskip
\mbox{}\\ \mbox
\bigskip

\begin{figure}[b]
\centering
\includegraphics[width=0.49\textwidth]{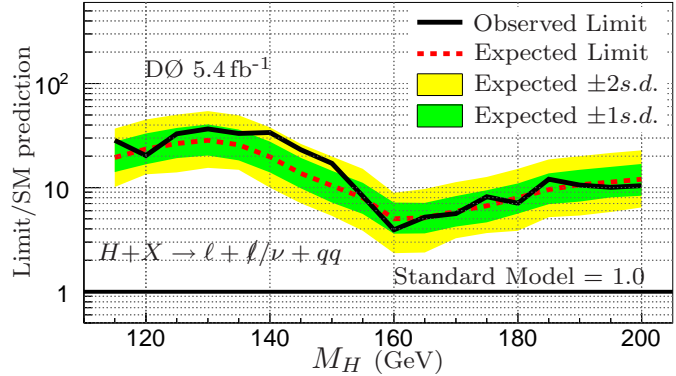}
\caption{\label{limits_combo}
Ratio of the 95\% CL limit to the SM expectation as a function of $M_H$ 
for the electron and muon channels combined.  
Expected limits are shown by the dashed line, observed limits are 
shown by the solid line.
The heavy and light shaded areas correspond to one and two standard deviations (s.d.)
around the expected result for the background-only hypothesis.
}
\end{figure}


\begin{thebibliography}{99}


\bibitem{Englert:1964et}F.~Englert and R.~Brout, Phys. Rev. Lett. {\bf 13}, 321 (1964).

\bibitem{PhysRevLett.13.508}P.~W. Higgs, Phys. Rev. Lett. {\bf 13}, 508 (1964).

\bibitem{Guralnik:1964eu}G.~S. Guralnik, C.~R. Hagen, and
T.~W.~B. Kibble, Phys. Rev. Lett. {\bf 13}, 585 (1964).

\bibitem{PhysRev.145.1156}P.~W. Higgs, Phys. Rev. {\bf 145}, 1156 (1966).

\bibitem{D0wwllnn} V.M.~Abazov {\sl et al.} (D0 Collaboration),
  Phys. Rev. Lett. {\bf 104}, 061804 (2010).

\bibitem{CDFwwllnn} T. Aaltonen {\sl et al.} (CDF Collaboration), 
  Phys. Rev. Lett. {\bf 104}, 061803 (2010).

\bibitem{HiggsCombo} T. Aaltonen, {\sl et al.} (CDF and D0 Collaborations), 
  Phys. Rev. Lett. {\bf 104}, 061802 (2010).

\bibitem{HiggsWidth} B. Dobrescu and J. Lykken, J. High Energy Phys. 
  {\bf 04}, 083 (2010).

\bibitem{GunionSoldate} J. F. Gunion and M. Soldate, 
  Phys. Rev. D {\bf 34}, 826 (1986).

\bibitem{Dzero} 
  V.M.~Abazov {\sl et al.} (D0 Collaboration),
  Nucl.\ Instrum.\ Methods in Phys. Res. A {\bf 565}, 463 (2006);
  M.~Abolins {\sl et al.}, Nucl. Instrum. and Methods A {\bf 584}, 75 (2007);
  R.~Angstadt {\sl et al.},  Nucl. Instrum. Methods Phys.
  Res., Sect. A {\bf 622}, 298 (2010).

\bibitem{jetalgo} 
  G.C. Blazey {\sl et al.}, arXiv:hep-ex/0005012 (2000). 

\bibitem{d0incjet}V.M.~Abazov {\sl et al.} (D0 Collaboration), 
Phys. Rev. Lett. {\bf 101}, 062001 (2008).

\bibitem{lumi} 
  T.~Andeen {\sl et al.}, FERMILAB-TM-2365 (2007).

\bibitem{alpgen} 
  M.L.~Mangano {\sl et al.},
  J. High Energy Phys. {\bf 07}, 001 (2003);
  version 2.11 was used.

\bibitem{Zpt}
  V.M.~Abazov {\sl et al.} (D0 Collaboration), Phys. Rev. Lett. {\bf 100}, 
  102002 (2008).

\bibitem{MelPet}
K.~Melnikov and F.~Petriello, Phys. Rev. D {\bf 74}, 114017 (2006).

\bibitem{comphep} 
  E.~Boos {\sl et al.} (CompHEP Collaboration),
  Nucl.\ Instrum.\ Methods in Phys. Res. A {\bf 534}, 250 (2004).

\bibitem{pythia}
  T.~Sj\"ostrand, S.~Mrenna, and P.~Skands,
  J. High Energy Phys. {\bf 05}, 026 (2006);
  version 6.409 with Tune~A was used.

\bibitem{cteq} 
  J.~Pumplin {\sl et al.},
  J. High Energy Phys. {\bf 07}, 012 (2002); 
  D.~Stump {\sl et al.},
  J. High Energy Phys {\bf 10}, 046 (2003).

\bibitem{mcfm} 
  J.M.~Campbell and R.K.~Ellis, 
  Phys.\ Rev.\ D {\bf 60}, 113006 (1999).

\bibitem{WZcross}
  R. Hamberg, W.L. van Neerven, and W.B. Kilgore, Nucl. Phys. B {\bf 359}, 343 
  (1991); {\bf 644}, 403(E) (2002).

\bibitem{MRST}
  A.D. Martin, R.G. Roberts, W.J. Stirling, and R.S. Thorne, Phys. Lett. 
  B{ \bf 604}, 61 (2004).

\bibitem{xsections} 
  M. Cacciari {\sl et al.}, J. High Energy Phys. {\bf 04}, 068 (2004);
  N. Kidonakis and R. Vogt, Phys. Rev. D {\bf 68}, 114014 (2003);
  N. Kidonakis, Phys. Rev. D {\bf 74}, 114012 (2006).


\bibitem{signal} 
  D. de Florian and M. Grazzini, Phys. Lett. B {\bf 674}, 291 (2009); 
  C. Anastasiou, R. Boughezal, and F. Petriello, J. High Energy Phys. {\bf 04}, 003 (2009); 
  E.L. Berger and J. Campbell, Phys. Rev. D {\bf 70}, 073011 (2004);
  T. Hahn {\sl et al.}, {\tt arXiv:hep-ph/0607308} (2006);
  M. L. Ciccolini, S. Dittmaier, and M. Kr\"amer, Phys. Rev. D {\bf 68}, 073003 (2003);
  O. Brein, A. Djouadi, and R. Harlander, Phys. Lett. B {\bf 579}, 149 (2004);
  J. Baglio and A. Djouadi, J. High Energy Phys. {\bf 10}, 064 (2010);
  A. Djouadi, J. Kalinowski, and M. Spira, Comput. Phys. Commun. 108, 56 (1998).


\bibitem{mc@nlo} S. Frixione and B.R. Webber, J. High Energy Phys. 
  {\bf 0206}, 029 (2002).

\bibitem{sherpa} T. Gleisberg {\sl et al.}, J. High Energy Phys. {\bf 02}, 
  056 (2004).

\bibitem{geant} 
  R. Brun and F. Carminati, CERN Program Library Long Writeup W5013, 1993 
  (unpublished).

\bibitem{d0tagprobe} V.M.~Abazov {\sl et al.} (D0 Collaboration), 
  Phys. Rev. D. {\bf 76}, 012003 (2007).

\bibitem{eta}
The pseudorapidity is defined as $\eta=-\ln[\tan(\theta/2)]$, 
where $\theta$ is the polar angle with respect to the proton beam direction.

\bibitem{d0sngltop} V.M.~Abazov {\sl et al.} (D0 Collaboration), 
  Phys. Rev. D. {\bf 78}, 012005 (2008).

\bibitem{d0sngltop0} V.M.~Abazov {\sl et al.} (D0 Collaboration), 
  Phys. Rev. D. {\bf 75}, 092007 (2007).

\bibitem{d0wmass} V.M.~Abazov {\sl et al.} (D0 Collaboration),
  Phys. Rev. Lett. {\bf 103}, 141801 (2009).

\bibitem{ttbarprd04} V.M.~Abazov {\sl et al.} (D0 Collaboration), 
  Phys. Rev. D. {\bf 74}, 112004 (2006).

\bibitem{Breiman} L. Breiman, Machine Learning {\bf 45}, 5-32 (2001).

\bibitem{Narsky} I. Narsky, arXiv:physics/0507143 [physics.data-an] (2005);
  I. Narsky, arXiv:physics/0507157 [physics.data-an] (2005).

\bibitem{herwig}
  G. Corcella {\sl et al.}, J. High Energy Phys. {\bf 01}, 010 (2001).

\bibitem{cls} 
  T. Junk, Nucl. Instrum. Methods in Phys. Res. A {\bf 434}, 435 (1999);
  A.~Read, J. Phys. G {\bf 28}, 2693 (2002).
 
\bibitem{wade} W. Fisher, FERMILAB-TM-2386-E.


\bibitem{appendix}
  Supplementary material is provided at \\ http://link.aps.org/supplemental/XX.YYYY/\\ PhysRevLett.XX.YYYYYY.

\end{thebibliography}
\end{document}